\documentclass[sigconf]{acmart}

\usepackage{algorithm}
\usepackage{algpseudocode}
\usepackage{multirow}
\usepackage{enumitem}
\usepackage{xcolor}
\usepackage{listings}
\usepackage{xcolor}

\definecolor{codegreen}{rgb}{0,0.6,0}
\definecolor{codegray}{rgb}{0.5,0.5,0.5}
\definecolor{codepurple}{rgb}{0.58,0,0.82}
\definecolor{backcolour}{rgb}{0.95,0.95,0.92}

\lstdefinestyle{mystyle}{
  backgroundcolor=\color{backcolour}, commentstyle=\color{codegreen},
  keywordstyle=\color{magenta},
  numberstyle=\tiny\color{codegray},
  stringstyle=\color{codepurple},
  basicstyle=\ttfamily\footnotesize,
  breakatwhitespace=false,         
  breaklines=true,                 
  captionpos=b,                    
  keepspaces=true,                 
  numbers=left,                    
  numbersep=5pt,                  
  showspaces=false,                
  showstringspaces=false,
  showtabs=false,                  
  tabsize=2
}

\definecolor{mygreen}{rgb}{0,0.6,0}
\definecolor{mygray}{rgb}{0.5,0.5,0.5}
\definecolor{mymauve}{rgb}{0.58,0,0.82}
\definecolor{myorange}{rgb}{1.0, 0.49, 0.0}

\usepackage[strict]{changepage}
\definecolor{darkblue}{rgb}{0.0, 0.0, 0.55}
\usepackage{framed}
\definecolor{formalshade}{rgb}{0.95,0.95,1}

\AtBeginDocument{%
  \providecommand\BibTeX{{%
    \normalfont B\kern-0.5em{\scshape i\kern-0.25em b}\kern-0.8em\TeX}}}

\acmConference[SIGIR '24]{The 47th International ACM SIGIR Conference on Research and Development in Information Retrieval}{July 14--18, 2024}{Washington D.C., USA}

\acmBooktitle{SIGIR '24: The 46th International ACM SIGIR Conference on Research and Development in Information Retrieval, July 14--18, 2024, Washington D.C., USA} 

\begin{document}

\title{OpenP5: An Open-Source Platform for Developing, Training, and Evaluating LLM-based Recommender Systems}

\author{Shuyuan Xu}
\affiliation{%
  \institution{Rutgers University}
  \country{New Brunswick, NJ, US}
}
\email{shuyuan.xu@rutgers.edu}

\author{Wenyue Hua}
\affiliation{%
  \institution{Rutgers University}
  \country{New Brunswick, NJ, US}
}
\email{wenyue.hua@rutgers.edu}

\author{Yongfeng Zhang}
\affiliation{%
  \institution{Rutgers University}
  \country{New Brunswick, NJ, US}
}
\email{yongfeng.zhang@rutgers.edu}

\renewcommand{\shortauthors}{Shuyuan Xu, Wenyue Hua and Yongfeng Zhang}

\begin{abstract}
In recent years, the integration of Large Language Models (LLMs) into recommender systems has garnered interest among both practitioners and researchers. Despite this interest, the field is still emerging, and the lack of open-source R\&D platforms may impede the exploration of LLM-based recommendations. This paper introduces OpenP5, an open-source platform designed as a resource to facilitate the development, training, and evaluation of LLM-based generative recommender systems for research purposes. The platform is implemented using encoder-decoder LLMs (e.g., T5) and decoder-only LLMs (e.g., Llama-2) across 10 widely recognized public datasets, catering to two fundamental recommendation tasks: sequential and straightforward recommendations. Recognizing the crucial role of item IDs in LLM-based recommendations, we have also incorporated three item indexing methods within the OpenP5 platform: random indexing, sequential indexing and collaborative indexing. Built on the Transformers library, the platform facilitates easy customization of LLM-based recommendations for users. OpenP5 boasts a range of features including extensible data processing, task-centric optimization, comprehensive datasets and checkpoints, efficient acceleration, and standardized evaluations, making it a valuable tool for the implementation and evaluation of LLM-based recommender systems. The open-source code and pre-trained checkpoints for the OpenP5 library are publicly available at \url{https://github.com/agiresearch/OpenP5}.

\end{abstract}





\maketitle

\section{Introduction}

The recent surge in interest around foundation models, including Large Language Models (LLM), within both academic and industrial domains has been largely attributed to their significant contributions across various research fields, including natural language processing (NLP) \cite{brown2020language, chung2022scaling, raffel2020exploring} and computer vision (CV) \cite{yuan2021florence, zhou2022learning}. In the sphere of recommender systems, practitioners and researchers are progressively incorporating these models into recommendation tasks. Certain recent studies, such as P5 \cite{p5} and M6 \cite{m6}, have efficaciously harnessed the advantages of large language models to facilitate generative recommendation by transforming recommendation tasks into natural language formats. Nevertheless, despite the intensifying focus on the utilization of foundation models within recommendation systems, the field remains relatively nascent, and the absence of standardized development platform might impede the rapid evolution of this budding area.

This paper endeavors to address the gap concerning the absence of standardized development platform in the realm of recommendation foundation models by introducing OpenP5. OpenP5 is an open-source platform for developing, training, and evaluating LLM-based models for generative recommendation, built upon the principles of the P5 model \cite{p5}. It incorporates four dimensions of the P5 model \cite{p5}: backbone models, downstream task, recommendation dataset, and item indexing method.

In recommender systems utilizing Large Language Models (LLMs), the generative prowess is derived from the foundational LLMs. Contemporary LLM architectures are principally classified into three types: encoder-only, encoder-decoder, and decoder-only. The burgeoning LLMs predominantly adopt either the encoder-decoder or the decoder-only architecture. To this end, the OpenP5 platform incorporates a quintessential LLM representative for both the encoder-decoder and decoder-only architectures. Specifically, the T5 \cite{raffel2020exploring} model is included as a representative of the encoder-decoder architecture, while the Llama-2 \cite{touvron2023llama} model epitomizes the decoder-only architecture.

In recommendation foundation models, language serves as an efficacious medium to integrate various recommendation downstream tasks into a singular model. Hence, OpenP5 considers the two most prevalent tasks in recommender systems: sequential recommendation and straightforward recommendation. The former requires the model to generate recommended items based on user ID and user history, while the latter mandates the model to generate recommendations solely based on user ID.

To facilitate researchers and practitioners, the OpenP5 platform includes a diverse range of commonly employed public recommendation datasets. 
We provide a comprehensive survey of the popular datasets used in recent years, and incorporates the top 10 datasets into the library.
We also design a Super P5 (SP5) model to have a preliminary exploration of the potential of recommendation foundation models that have the ability to recommend items across various datasets
using a singular model. 

In OpenP5, we also include various methods to represent items in language. The cruciality of assigning a unique ID to each item in recommendation foundation models is underscored, ensuring that each item is represented by a minimal number of tokens and can be differentiated from other items, and to avoid hallucination problems in generative recommendation \cite{hua2023index}. Moreover, the item indexing method can greatly impact the performance of the recommendation foundation models. Existing studies have adapted several item representation methods while transforming recommendation tasks into language generation tasks. For instance, P5 \cite{p5} uses number tokens, M6 \cite{m6} leverages rich metadata to generate metadata-based embeddings to represent items, and LMRecSys \cite{Zhang2021language} utilizes item titles as representation. However, considering many public datasets may not include rich metadata or textual information, OpenP5 platform includes only three item indexing methods based solely on user-item interactions: random indexing, sequential indexing, and collaborative indexing \cite{hua2023index}.

In summary, OpenP5 offers a platform for developing, training and evaluating LLM-based recommendation systems 
based on the P5 principles \cite{p5}, encompassing two downstream tasks, ten datasets, three ID creation methods, and supporting both encoder-decoder and decoder-only LLM architectures. It also provides checkpoints based on two backbone models for each of the ten popular public datasets and an implementation of SP5 pre-trained on all datasets over three item indexing methods. Furthermore, the platform also supports users to develop their customized methods based on our provided APIs, such as new ID creation methods, backbones, datasets, tasks, or evaluation methods, facilitating future research on LLM-based generative recommendation.

The remainder of this paper is organized as follows. In Section \ref{sec:related}, we provide the necessary background and related work.
In Section \ref{sec:dataprocessing} to Section \ref{sec:experiments}, we introduce how to process data in OpenP5, including raw data preprocessing, item indexing methods, personalized prompt collection and data preparation, explain the pre-training and fine-tuning details of OpenP5, provide the evaluation methods, and show the experimental results, which can help users of the platform to easily adapt OpenP5 platform to other data or tasks.
Finally, we conclude the work and discuss the future directions in Section \ref{sec:conclusion}.

\section{Related Work}\label{sec:related}

Recently, there have been several attempts to leverage the power of large language models into recommender systems. Following \cite{hua2023tutorial, lin2023can}, we introduce existing work in three dimensions: the role of LLM, how to adapt LLM, and the recommendation tasks.

With the powerful ability, LLM can participate in several components of recommender systems. LLM can be used for feature engineering, which takes the original data as input and generates rich textual features as data augmentations \cite{xi2023towards,mysore2023large}. LLM can also be used as feature representation extractor, which formulates features as embeddings. Using LLM to obtain feature representation can provide item or user representations with rich semantic information \cite{yu2021tiny,li2023exploring,wu2022mm} and may be helpful for cross-domain or cold-start recommendation with natural language as connections \cite{ding2021zero,wang2022transrec,fu2023exploring}. Some works \cite{p5,m6,li2023pbnr,liu2023pre} directly use LLM as recommender system, which is able to complete recommendation related tasks by LLM. In addition to being part of the recommendation system, LLM can also be used as a controller, possibly leading to a more interactive and explainable recommendation \cite{gao2023chat,friedman2023leveraging}.

Regarding how to adapt LLM, the LLM in recommender systems can be either tuned or not tuned, depending on whether the model will tune LLM during the training section or not. This includes both full fine-tuning and other parameter-efficient fine-tuning methods, such as LoRA \cite{hu2021lora}. With the emergence of large foundation models, researchers intend to analyze the zero-shot or few-shot performance of LLM in recommendation scenarios. Many existing work \cite{sileo2022zero,liu2023chatgpt,wang2023zero,sun2023chatgpt,dai2023uncovering,li2023preliminary} investigate zero-shot recommendation based on LLM without fine-tuning, which constructs prompts to instruct LLM on various recommendation tasks, such as rating prediction, pairwise comparison, reranking, etc. Although LLM may provide good language understanding performance, the recommendation performance without fine-tuning still needs future improvement, which indicates the importance of domain knowledge such as user-item interaction information from recommender systems \cite{liu2023chatgpt,dai2023uncovering}. 

With the development of the field of recommender systems, the tasks of recommender systems are not limited to rating prediction or item recommendation. Traditional recommender systems are usually designed for a specific task, due to the difference on data format and model architecture for different tasks. With the help of LLM, language can serve as a bridge to integrate various downstream tasks into a singular model. Existing work on LLM-based recommender systems can be divided into multi-task recommender and task-spcific recommender from the perspective of downstream tasks. Some work \cite{ji2023genrec,petrov2023generative,li2023personalized} incorporates LLM to improve the performance on specific tasks, while some other work \cite{p5,m6,zhang2023recommendation} uses LLM to handle multi-task through a unified language format.

Based on the aspects introduced above, we define the OpenP5 platform as a tool for researchers and practitioners who are interested in developing fine-tuned LLM as recommender systems to perform multiple recommendation tasks.


\begin{figure}[t]
    \centering
    \includegraphics[width=0.4\textwidth]{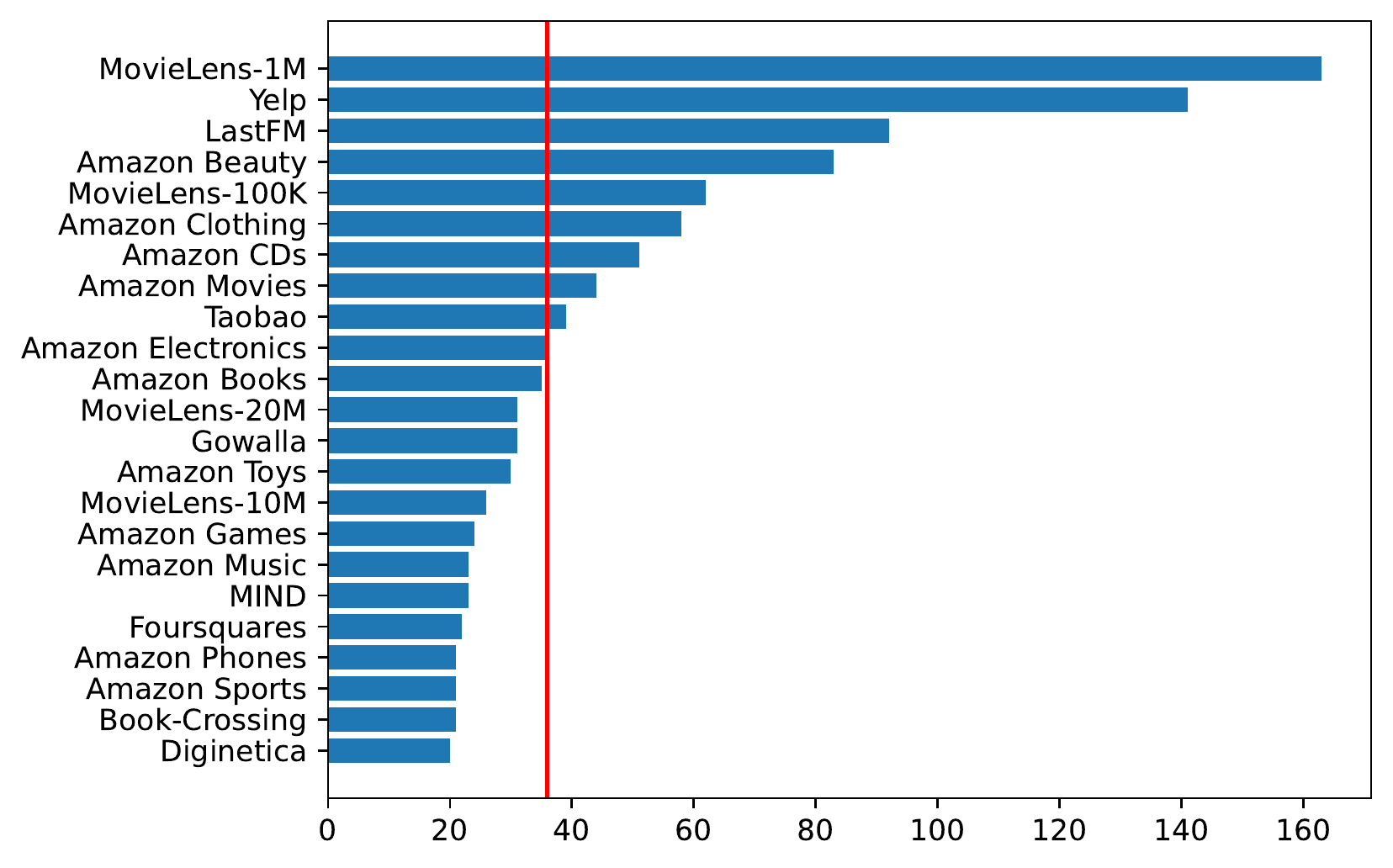}
    \vspace{-15pt}
    \caption{The frequency of public datasets in recent publications. We only show the datasets with more than 20 occurrences in recent three years at SIGIR, RecSys, WSDM, KDD, WWW, and CIKM. We select top 10 datasets in OpenP5 library. The vertical red line represents the threshold.}
    \label{fig:datasetfrequency}
    \vspace{-10pt}
\end{figure}

\begin{table*}[t]
\small
    \begin{tabular}{ll}
    \toprule
    User & Interaction History \\\midrule
    A1YJEY40YUW4SE & B004756YJA B004ZT0SSG B0020YLEYK 7806397051 B002WLWX82 \\
A60XNB876KYML & B0009P4PZC B009HULFLW B00BZ1QN2C B00G2TQNZ4 B00812ZWOS 7806397051 B0000YUX4O \\
A3G6XNM240RMWA & 7806397051 B003H8180I B00538TSMU B002S8TOYU B001MP471K B00011JI88 B00C1F13CQ B003ATNYJC B003ZS6ONQ \\
A1PQFP6SAJ6D80 & B0030HKJ8I B00027D8IC B002PMLGOU B00BN1MPPS 7806397051 B004Z40048\\
A38FVHZTNQ271F & 7806397051 B009PZVOF6 B008LQX8J0 B007EHWDTS B009DDGHFC B002BGDLDO B003VWZCMK B00DQ2ILQY B00DAYGJVW \\\bottomrule 
    \end{tabular}
    \caption{This table presents a few instances from \textit{Amazon Beauty} data. The data is stored a text file format, with each line encapsulating the information pertaining to an individual user. Within every line, the initial element represents the user's raw ID, followed by item raw IDs, listed in chronological order based on the user's interaction history.}
    \label{tab:data}
    \vspace{-10pt}
\end{table*}

\section{Data Processing}\label{sec:dataprocessing}
In this section, we will discuss the data processing module of our platform, which allows users to integrate new datasets and create customized extensions.
\subsection{Raw Data Preprocessing}
The OpenP5 platform provides 10 popular preprocessed public datasets. To identify popular public datasets suitable for recommendations, we conduct a frequency analysis of their occurrence in recent publications. More specifically, we examine papers accepted in the preceding three years at related conferences, including SIGIR, RecSys, WSDM, KDD, WWW, and CIKM. Using the ACM Digital Library\footnote{https://dl.acm.org/}, we filter relevant publications with the keywords ``recommend'', ``recommender'', ``recommendation'' and ``collaborative''. Due to the large number of public datasets, we only show the frequency of datasets with more than 20 occurrences in Figure \ref{fig:datasetfrequency}. We include the top 10 popular public datasets in the OpenP5 library, including \textit{Movielens-1M}, \textit{Yelp}, \textit{LastFM}, \textit{Amazon Beauty}, \textit{Movielens-100K}, \textit{Amazon Clothing}, \textit{Amazon CDs}, \textit{Amazon Movies}, \textit{Taobao}, and \textit{Amazon Electronics}. This approach ensures that the datasets selected are not only popular but also align with current research trends in recommender systems. We provide the statistical overview of all datasets in our GitHub repository\footnote{https://github.com/agiresearch/OpenP5}.


The preprocessed data is saved as a txt file, with an illustrative example from the \textit{Amazon Beauty} dataset shown in Table \ref{tab:data}. Our platform principally requires user-item interaction data, as additional information may not be available for most public datasets. More precisely, we segregate the information of different users into separate lines. Within each line, elements are divided by a space, where the first element denotes the user raw ID, and the subsequent elements -- item raw IDs -- chronologically delineate the user's interaction history. Users of the platform can effortlessly train models on new datasets by converting the raw data into the specified data format. The platform will automatically partition the data into training, validation, and testing sets.

\subsection{Item Indexing}

In order to transform recommendation tasks into language generation tasks, user and item identifiers need to be compatible with natural language. This compatibility ensures that these identifiers can be seamlessly incorporated into natural language instructions used for the pre-training, fine-tuning, and prompting stages of Large Language Models (LLMs). Our platform provides implementation of three item indexing methods: random indexing, sequential indexing, and collaborative indexing. After applying indexing methods, the results are saved as a txt file consisting of two values in each line, where the first value represents the raw ID, and the second value represents the reindexed ID. The user-item interaction data will be formulated in the same format as preprocessed data (i.e., Table \ref{tab:data}) after indexing. We will introduce more details about the provided indexing methods in the platform.

\textbf{Random Indexing.} Random indexing represents a straightforward approach to item indexing. This method assigns each item a unique random number that serves as the item ID. Within the model, the SentencePiece tokenizer \cite{sennrich2016neural} is employed to further tokenize this random number ID into a sequence of tokens. For instance, an item with the randomly assigned unique ID of ``2048'' would be tokenized into the tokens ``20'' and ``48'' within the recommendation foundation model. 

While random indexing is frequently employed in traditional recommendation systems, it may not be optimally suited for foundation models \cite{hua2023index}: the potential drawback stems from the fact that the randomly assigned IDs are further tokenized, which can inadvertently cause unrelated items to share identical tokens. To illustrate, the items ``2048'' and ``2049'', despite being completely unrelated and not even interacted with by the same user, share the token ``20''. Consequently, the model could mistakenly establish a semantic relationship between these items. As the relationship stems from the index structure, they are impossible to eliminate no matter how the model learns from data, thereby affecting the accuracy of the recommendations \cite{hua2023index}. Consequently, RID is considered an unfavorable method. However, we still include this simple indexing method in this platform in case researchers would like to use it as a baseline for comparison and exploration.

\textbf{Sequential Indexing.} To mitigate the issues associated with random indexing, one viable strategy involves integrating collaborative information into item IDs. A basic implementation of this approach can be observed in the sequential indexing method, as illustrated in \cite{hua2023index}. This method assigns consecutive number IDs to users' consecutive interactions, commencing with the first user and progressing through to the last user. It iterates across all interactions, assigning a new, incrementally increasing ID to any item that has not yet been assigned. Importantly, we apply the sequential indexing method solely to the training data to circumvent potential data leakage during the evaluation phase.

For items subjected to sequential indexing, the sharing of an identical token at the same position following tokenization between two items suggests that these two items may have been interacted with by the same user. Consequently, the sequential information embedded within the item IDs could potentially augment the effectiveness of the foundational model's recommendations.

\begin{algorithm}[t]
\caption{Method for Collaborative Indexing}\label{alg:collaborative}
\begin{algorithmic}[1]
\Require Training data user sequence $D$, number of clusters $N$ to be created, number of items $k$ in the largest allowed cluster
\State Instantiate a queue and enqueue all items as one set
\While{queue is not empty}
\State Dequeue the first item set $S$
\If{The size of $S$ $<$ $k$}
\State Assign a unique token to all items within $S$
\Else
\State Compute the co-occurrence matrix $M$ for items in $S$ based on $D$
\State Apply spectral clustering on $M$ into $N$ clusters
\State Generate unique tokens for each cluster and assign corresponding tokens to all items within $S$ based on the clustering
\State Enqueue all resultant clusters
\EndIf
\EndWhile
\end{algorithmic}
\end{algorithm}

\textbf{Collaborative Indexing.} To integrate further collaborative information into item indexing, we have incorporated a collaborative indexing method within the OpenP5 library. The underlying intuition of the collaborative indexing method is predicated on the idea that the frequency of co-occurrence of items should influence the degree to which they share the same token at the same position \cite{hua2023index}. This concept is represented as a graph, with nodes signifying items and edge weights denoting co-occurrence frequency. To engender collaborative indexing, we employ the spectral clustering method \cite{ng2001spectral,von2007tutorial}.
Given that the collaborative indexing method necessitates the introduction of Out-of-vocabulary (OOV) tokens to construct item indices, we denote these OOV tokens with angle brackets "$\langle\rangle$" (e.g., "$\langle CI1\rangle$"). A detailed exposition of this method can be found in Algorithm \ref{alg:collaborative}.

\subsection{Personalized Prompt Collection}

Recommendation foundation models possess the capability to integrate various downstream tasks of recommendation into a singular generative model \cite{p5,m6}. Acknowledging that some public datasets may not encompass certain information such as reviews, metadata, explicit feedback, and so forth, the OpenP5 library focuses solely on the two most commonly employed downstream tasks in recommender systems: the \textbf{sequential recommendation} task and the \textbf{straightforward recommendation} task. Both of the tasks encompass several personalized prompts tailored to individual users. More specifically, various prompt templates have been designed for both tasks, which are filled with personalized information such as user ID and item ID. In addition, to circumvent the issue of recommending items from divergent datasets in SP5 (for instance, recommending a Yelp restaurant to an Amazon user), the dataset name has been included within our designed prompts.

The sequential recommendation task requires generating recommended items based on the user's history, and thus the personalized prompts incorporate the dataset name, user ID, user history, and target item ID. The straightforward recommendation task requires the model to generate recommended items given only the user ID, hence the prompts for this task exclude user history.

The following examples illustrate the prompt templates for both downstream tasks.

\noindent\textbf{Sequential Recommendation}\\
\texttt{Input Template: Considering \{\textcolor{blue}{dataset}\} user\_\{\textcolor{blue}{user\_id}\} has interacted with \{\textcolor{blue}{dataset}\} items \{\textcolor{blue}{history}\} . What is the next recommendation for the user ?\\
Target Template: \{\textcolor{blue}{dataset}\} \{\textcolor{blue}{target}\}
}

\noindent\textbf{Straightforward Recommendation}\\
\texttt{Input Template: What should we recommend for \{\textcolor{blue}{dataset}\} user\_\{\textcolor{blue}{user\_id}\} ?\\
Target Template: \{\textcolor{blue}{dataset}\} \{\textcolor{blue}{target}\}
}

\begin{table*}[t]
\centering
    \small
    \begin{tabular}{llp{12.1cm}l}
    \toprule
    Task & Type & Input & Output \\\midrule
    sequential & seen & What would \{dataset\} user\_\{user\_id\} be likely to purchase next after buying \{dataset\} items \{history\} ? & \{dataset\} \{target\} \\
    sequential & unseen & What is the top recommended item for \{dataset\} user\_\{user\_id\} who interacted with \{dataset\} item \{history\} ? & \{dataset\} \{target\}\\
    straightforward & seen & What should we recommend for \{dataset\} user\_\{user\_id\} ? & \{dataset\} \{target\}\\
    straightforward & unseen & What is the top recommendation for \{dataset\} user\_\{user\_id\} ? & \{dataset\} \{target\} \\\bottomrule
    \end{tabular}
    
    \caption{This table displays a few instances of prompt templates. The prompt templates are stored in a text file, with each line representing a unique prompt template. Every prompt encompasses four types of information, delineated by semicolons: the recommendation task, indication of whether seen or unseen during training, input template, and output template. }
    \vspace{-10pt}
    \label{tab:prompt}
\end{table*}


The OpenP5 platform offers 11 distinct prompt templates for each downstream task. From each task, a single prompt template is selected as the \textit{unseen} prompt, which serves to evaluate the model's zero-shot generalization capabilities. Notably, the OpenP5 platform is designed with flexibility in mind, enabling users to modify the prompt templates according to their specific needs or objectives. More specifically, the prompt templates are saved in a txt file, with a representative example depicted in Table \ref{tab:prompt}. Each line delineates a distinct prompt template, encompassing four pieces of information separated by semicolons: the first specifies the task to which the prompt pertains; the second denotes whether this prompt template is exposed to the model during training; the third outlines the instruction for input; and the fourth signifies the recommended item as output. The personalized information within the prompt template is enclosed in curly brace, and are substituted with specific data during data processing.

\section{Multi-task Learning}
In our previous discussion, we highlighted that the OpenP5 platform supports the two predominant recommendation tasks: sequential and straightforward recommendations. It worth mentioning that the platform is also architecturally equipped to incorporate additional tasks into its training paradigm. This extensibility is a cornerstone of the platform's design, allowing for a broader scope of learning and adaptability. This section is devoted to elucidating the multi-task learning framework provided by the platform.

When tasks are learned in a sequence rather than concurrently, the model is susceptible to the ``forgetting problem'' \cite{kemker2018measuring,kirkpatrick2017overcoming,li2017learning,yuan2021one}, which predominantly enhances its performance on the latest task to the detriment of prior tasks. To counteract this, simultaneous task learning is imperative. A common, albeit intuitive, solution is to mix training data from various tasks. This method, however, is not without its pitfalls. In the case of our Super P5 (SP5) model, indiscriminately blending data from disparate datasets can still lead to the ``forgetting problem'' when the datasets are uneven in size. Additionally, tasks differ in their textual length requirements; sequential recommendations necessitate a history of interactions, resulting in longer input sequences compared to direct recommendations. This discrepancy can lead to excessive padding when batching data from multiple tasks. To avoid these issues, our platform ensures that each batch is task-homogeneous—data from the same task. 
This strategy effectively mitigates forgetting and maintains efficiency across varying task demands.

\section{Pre-training and Fine-tuning}
Given the personalized prompt for multiple recommendation tasks, we then introduce the pre-training and fine-tuning of the OpenP5 platform.

To improve the efficiency of pre-training and fine-tuning, the OpenP5 platform incorporates two techniques. One is to enable distributed learning in Multi-GPU environment. Distributed learning enables the model to be learned within a shorter time to improve the efficiency. Apart from distributed learning, the OpenP5 platform incorporates efficient training methods, such as LoRA \cite{hu2021lora}, which freezes the pre-trained model weights and injects trainable rank decomposition matrices to reduce the trainable parameters and improve the efficiency.


\section{How to Customize OpenP5}
The platform facilitates the development of customized LLM-based recommendation models by users. This section illustrates how OpenP5 can be adapted from various angles through illustrative examples.

\begin{itemize}[leftmargin=*]
    \item \textbf{Incorporating New Datasets:} Integrating new datasets into the recommendation model is straightforward with our platform, provided that the datasets are properly formatted. This ease of integration supports the seamless training of LLM-based recommendation models with new data sources.
    
    \item \textbf{User/Item ID:} Our platform supports three distinct methods for indexing items, with the flexibility to introduce additional indexing strategies. For instance, adopting item titles as unique identifiers can be achieved by generating preprocessed data that utilizes these new IDs.
    
    \item \textbf{Adopting New Backbone Models:} Given that the platform's architecture is predicated on Transformers library\footnote{\url{https://huggingface.co/docs/transformers/en/index}}, users can conveniently replace backbone models with alternative ones from the Transformers library. 
    
    \item \textbf{Customizing Personalized Prompts:} Personalized prompt templates on the platform can be readily replaced, allowing for the inclusion of novel tasks into the training process. Moreover, the platform's capability to manage out-of-vocabulary (OOV) tokens enhances its utility for LLM-based recommendation models that require such functionality. For example, \citet{geng2023vip5} introduced a multimodal foundation model that integrates item images into prompts, utilizing a visual encoder to transform images into image tokens, which are treated as OOV by the tokenizer.
\end{itemize}

In summary, OpenP5 exhibits remarkable flexibility in accommodating new datasets, indexing methodologies, backbone models, and tasks. This adaptability underscores its potential as a foundational tool for pioneering research in the domain of LLM-based generative recommendation systems.

\begin{table*}[t]
    \centering
    \small
    \setlength{\tabcolsep}{1.5pt}
    \begin{tabular}{ccccccccccccc}
    \toprule
        \multirow{2}{*}{Methods} & \multicolumn{4}{c}{\textbf{ML1M}} & \multicolumn{4}{c}{\textbf{Beauty}} & \multicolumn{4}{c}{\textbf{LastFM}}\\
        \cmidrule(lr){2-5} \cmidrule(lr){6-9} \cmidrule(lr){10-13}
         & HR@5 & NCDG@5 & HR@10 & NCDG@10 & HR@5 & NCDG@5 & HR@10 & NCDG@10 & HR@5 & NCDG@5 & HR@10 & NCDG@10 \\
        \midrule
        Caser & 0.0912 & 0.0565 & 0.1442 & 0.0734 & 0.0205 & 0.0131 & 0.0347 & 0.0176 & 0.0303 & 0.0178 & 0.0413 & 0.0214 \\
        HGN & 0.1430 & 0.0874 & 0.2404 &  0.1231  & 0.0325 & 0.0206 & 0.0512 & 0.0266 & 0.0321 & 0.0175 & 0.0505 & 0.0233 \\
        GRU4Rec & 0.0806 & 0.0475 & 0.1344 & 0.0649 & 0.0164 & 0.0099 & 0.0283 & 0.0137 & 0.0275 & 0.0158 &  0.0367 & 0.0187 \\
        BERT4Rec & 0.1308 & 0.0804 & 0.2219 & 0.1097 & 0.0203 & 0.0124 & 0.0347 & 0.0170 & 0.0422 & 0.0269 & 0.0633 & 0.0337  \\
        FDSA & 0.1167 & 0.0762 & 0.1868 & 0.0987 & 0.0267 & 0.0163 & 0.0407 & 0.0208 & 0.0303 & 0.0219 & 0.0413 & 0.0254 \\
        SASRec & 0.1078 & 0.0681 & 0.1810 & 0.0918 & 0.0387 & 0.0249 & 0.0605 & 0.0318 & \textbf{0.0505} & \underline{0.0331} & \underline{0.0688} & \underline{0.0390} \\
        \midrule
        OpenP5-T5-R (seen) & 0.1098 & 0.0734 & 0.1575 & 0.0888 & 0.0318 & 0.0226 & 0.0464 & 0.0273 & 0.0156 & 0.0104 & 0.0312 & 0.0153 \\
        OpenP5-T5-S (seen) & 0.1901 & 0.1229 & 0.2849 & 0.1535 & \textbf{0.0457} & \textbf{0.0336} & \textbf{0.0622} & \textbf{0.0389} & 0.0394 & 0.0262 & 0.0578 & 0.0321\\
        OpenP5-T5-C (seen) & \textbf{0.2066} & \textbf{0.1400} & \textbf{0.2945} & \underline{0.1683} & 0.0421 & 0.0285 & 0.0601 & 0.0346 & 0.0453 & 0.0301 & 0.0674 & 0.0370\\
        SP5-T5-R (seen) & 0.0305 & 0.0185 & 0.0558 & 0.0267 & 0.0073 & 0.0050 & 0.0111 & 0.0062 & 0.0037 & 0.0022 & 0.0064 & 0.0031\\
        SP5-T5-S (seen) & 0.1058 & 0.0696 & 0.1589 & 0.0866 & 0.0130 & 0.0067 & 0.0224 & 0.0097 & 0.0101 & 0.0061 & 0.0156 & 0.0079 \\
        SP5-T5-C (seen) & 0.1490 & 0.0984 & 0.2225 & 0.1221 & 0.0276 & 0.0192 & 0.0391 & 0.0229 & 0.0192 & 0.0130 & 0.0284 & 0.0160 \\
        OpenP5-Llama-R (seen) & 0.0300 & 0.0197 & 0.0470 & 0.0252 & 0.0018 & 0.0013 & 0.0024 & 0.0015 & 0.0193 & 0.0120 & 0.0284 & 0.0149\\
        OpenP5-Llama-S (seen) & 0.0714 & 0.0466 & 0.1094 & 0.0587 & 0.0022 & 0.0036 & 0.0013 & 0.0017 & 0.0101 & 0.0059 & 0.0202 & 0.0092\\
        OpenP5-Llama-C (seen) & 0.0012 & 0.0006 & 0.0026 & 0.0011 & 0.0002 & 0.0001 & 0.0007 & 0.0003 & 0.0018 & 0.0013 & 0.0018 & 0.0013\\
        SP5-Llama-R (seen) & 0.0008 & 0.0004 & 0.0033 & 0.0012 & 0.0007 & 0.0004 & 0.0014 & 0.0006 & 0.0028 & 0.0017 & 0.0055 & 0.0026\\
        SP5-Llama-S (seen) & 0.0045 & 0.0026 & 0.0118 & 0.0049 & 0.0009 & 0.0004 & 0.0017 & 0.0007 & 0.0009 & 0.0005 & 0.0037 & 0.0014\\
        SP5-Llama-C (seen) & 0.0026 & 0.0015 & 0.0041 & 0.0019 & 0.0004 & 0.0002 & 0.0009 & 0.0004 & 0.0009 & 0.0004 & 0.0018 & 0.0007\\
        \midrule
        OpenP5-T5-R (unseen) & 0.1058 & 0.0693 & 0.1533 & 0.0847 & 0.0313 & 0.0222 & 0.0456 & 0.0267 & 0.0128 & 0.0072 & 0.0248 & 0.0110 \\
        OpenP5-T5-S (unseen) & 0.1916 & 0.1236 & 0.2854 & \textbf{0.1737} & \underline{0.0452} & \underline{0.0332} & \underline{0.0613} & \underline{0.0384} & 0.0404 & 0.0265 & 0.0606 & 0.0331\\
        OpenP5-T5-C (unseen) & \underline{0.2055} & \underline{0.1386} & \underline{0.2940} & 0.1672 & 0.0412 & 0.0286 & 0.0600 & 0.0346 & \underline{0.0504} & \textbf{0.0332} & \textbf{0.0724} & \textbf{0.0420}\\
        SP5-T5-R (unseen) & 0.0306 & 0.0190 & 0.0541 & 0.0264 & 0.0076 & 0.0051 & 0.0116 & 0.0064 & 0.0046 & 0.0025 & 0.0092 & 0.0039\\
        SP5-T5-S (unseen) & 0.1046 & 0.0688 & 0.1586 & 0.0862 & 0.0183 & 0.0122 & 0.0266 & 0.0149 & 0.0083 & 0.0049 & 0.0147 & 0.0070\\
        SP5-T5-C (unseen) & 0.1064 & 0.0702 & 0.1685 & 0.0901 & 0.0240 & 0.0162 & 0.0339 & 0.0193 & 0.0101 & 0.0080 & 0.0211 & 0.0117 \\
        OpenP5-Llama-R (unseen) & 0.0296 & 0.0200 & 0.0444 & 0.0247 & 0.0017 & 0.0011 & 0.0022 & 0.0013 & 0.0183 & 0.0108 & 0.0202 & 0.0113\\
        OpenP5-Llama-S (unseen)  & 0.0556 & 0.0364 & 0.0877 & 0.0467 & 0.0029 & 0.0017 & 0.0045 & 0.0022 & 0.0128 & 0.0078 & 0.0202 & 0.0103\\
        OpenP5-Llama-C (unseen)  & 0.0010 & 0.0006 & 0.0018 & 0.0009 & 0.0004 & 0.0002 & 0.0007 & 0.0003 & 0.0009 & 0.0004 & 0.0046 & 0.0015\\
        SP5-Llama-R (unseen) & 0.0015 & 0.0009 & 0.0028 & 0.0013 & 0.0013 & 0.0007 & 0.0019 & 0.0010 & 0.0009 & 0.0006 & 0.0028 & 0.0012\\
        SP5-Llama-S (unseen) & 0.0048 & 0.0030 & 0.0106 & 0.0048 & 0.0008 & 0.0005 & 0.0013 & 0.0006 & 0.0028 & 0.0018 & 0.0046 & 0.0024\\
        SP5-Llama-C (unseen) & 0.0026 & 0.0016 & 0.0043 & 0.0021 & 0.0004 & 0.0002 & 0.0019 & 0.0004 & 0.0009 & 0.0005 & 0.0009 & 0.0005\\
        
        \bottomrule
    \end{tabular}
    \caption{Performance results on sequential recommendation task. R, S, C represent three item indexing.}
    \label{tab:sequential}
    \vspace{-20pt}
\end{table*}

\begin{table*}[t]
    \centering
    \small
    \setlength{\tabcolsep}{1.5pt}
    \begin{tabular}{cccccccccccccccc}
    \toprule
        \multirow{2}{*}{Methods} & \multicolumn{4}{c}{\textbf{ML1M}} & \multicolumn{4}{c}{\textbf{Beauty}} & \multicolumn{4}{c}{\textbf{LastFM}}\\
        \cmidrule(lr){2-5} \cmidrule(lr){6-9} \cmidrule(lr){10-13}
         & HR@5 & NCDG@5 & HR@10 & NCDG@10 & HR@5 & NCDG@5 & HR@10 & NCDG@10 & HR@5 & NCDG@5 & HR@10 & NCDG@10 \\
        \midrule
        BPR-MF & 0.0141 & 0.0081 & 0.0301 & 0.0133 & 0.0224 & 0.0149 & 0.0363 & 0.0204 & 0.0218 & 0.0147 & 0.0253 & 0.0162\\
        BPR-MLP & 0.0123 & 0.0068 & 0.0270 & 0.0116 & 0.0193 &  0.0127 & 0.0305 & 0.0176 & 0.0211 & 0.0150 & 0.0321 & 0.0185\\
        SimpleX & 0.0301 & 0.0133 & \underline{0.0596} & 0.0206 & \underline{0.0300} & 0.0189 & \textbf{0.0471} & 0.0245 & 0.0312 & 0.0211 & 0.0523 & 0.0277\\
        \midrule
        OpenP5-T5-R (seen) & 0.0215 & 0.0133 & 0.0348 & 0.0176 & 0.0233 & 0.0169 & 0.0317 & 0.0196 & 0.0239 & 0.0151 & 0.0294 & 0.0169\\
        OpenP5-T5-S (seen) & \underline{0.0310} & \underline{0.0192} & 0.0571 & \underline{0.0275} & \textbf{0.0317} & \textbf{0.0239} & 0.0437 & \textbf{0.0277} & \underline{0.0376} & \underline{0.0259} & \textbf{0.0661} & \textbf{0.0350}\\
        OpenP5-T5-C (seen) & \textbf{0.0347} & \textbf{0.0224} & \textbf{0.0618} & \textbf{0.0309} & 0.0294 & \underline{0.0206} & \underline{0.0444} & \underline{0.0254} & \textbf{0.0404} & \textbf{0.0270} & \underline{0.0615} & \underline{0.0336}\\
        SP5-T5-R (seen) & 0.0114 & 0.0074 & 0.0243 & 0.0115 & 0.0039 & 0.0021 & 0.0087 & 0.0036 & 0.0012 & 0.0009 & 0.0073 & 0.0022\\
        SP5-T5-S (seen) & 0.0121 & 0.0082 & 0.0224 & 0.0115 & 0.0130 & 0.0067 & 0.0224 & 0.0097 & 0.0027 & 0.0012 & 0.0073 & 0.0027 \\
        SP5-T5-C (seen) & 0.0214 & 0.0135 & 0.0344 & 0.0177 & 0.0176 & 0.0121 & 0.0278 & 0.0154 & 0.0183 & 0.0118 & 0.0321 & 0.0161\\
        OpenP5-Llama-R (seen) & 0.0106 & 0.0061 & 0.0205 & 0.0093 & 0.0017 & 0.0012 & 0.0023 & 0.0014 & 0.0202 & 0.0122 & 0.0275 & 0.0146\\
        OpenP5-Llama-S (seen) & 0.0103 & 0.0066 & 0.0210 & 0.0104 & 0.0050 & 0.0035 & 0.0065 & 0.0040 & 0.0147 & 0.0112 & 0.0220 & 0.0134\\
        OpenP5-Llama-C (seen) & 0.0012 & 0.0007 & 0.0022 & 0.0011 & 0.0002 & 0.0001 & 0.0003 & 0.0002 & 0.0028 & 0.0046 & 0.0022 & 0.0028 \\
        SP5-Llama-R (seen) & 0.0018 & 0.0008 & 0.0041 & 0.0016 & 0.0007 & 0.0004 & 0.0014 & 0.0006 & 0.0018 & 0.0018 & 0.0028 & 0.0021\\
        SP5-Llama-S (seen) & 0.0063 & 0.0034 & 0.0119 & 0.0052 & 0.0007 & 0.0004 & 0.0016 & 0.0007 & 0.0009 & 0.0004 & 0.0028 & 0.0009\\
        SP5-Llama-C (seen) & 0.0022 & 0.0012 & 0.0041 & 0.0018 & 0.0002 & 0.0001 & 0.0008 & 0.0003 & 0.0001 & 0.0001 & 0.0018 & 0.0006\\
        \midrule
        OpenP5-T5-R (unseen) & 0.0177 & 0.0114 & 0.0301 & 0.0154 & 0.0078 & 0.0051 & 0.0127 & 0.0066 & 0.0183 & 0.0117 & 0.0284 & 0.0150\\
        OpenP5-T5-S (unseen) & 0.0190 & 0.0122 & 0.0368 & 0.0178 & 0.0122 & 0.0089 & 0.0182 & 0.0109 & 0.0128 & 0.0076 & 0.0202 & 0.0099\\
        OpenP5-T5-C (unseen) & 0.0210 & 0.0134 & 0.0303 & 0.0164 & 0.0139 & 0.0089 & 0.0226 & 0.0117 & 0.0174 & 0.0117 & 0.0257 & 0.0144 \\
        SP5-T5-R (unseen) & 0.0086 & 0.0056 & 0.0209 & 0.0095 & 0.0017 & 0.0009 & 0.0042 & 0.0017 & 0.0018 & 0.0009 & 0.0046 & 0.0017 \\
        SP5-T5-S (unseen)  & 0.0105 & 0.0066 & 0.0177 & 0.0088 & 0.0044 & 0.0024 & 0.0093 & 0.0040 & 0.0073 & 0.0039 & 0.0147 & 0.0062 \\
        SP5-T5-C (unseen) & 0.0126 & 0.0082 & 0.0238 & 0.0117 & 0.0068 & 0.0034 & 0.0113 & 0.0049 & 0.0028 & 0.0011 & 0.0092 & 0.0032\\
        OpenP5-Llama-R (unseen) & 0.0094 & 0.0063 & 0.0190 & 0.0094 & 0.0014 & 0.0021 & 0.0011 & 0.0013 & 0.0202 & 0.0139 & 0.0202 & 0.0139\\
        OpenP5-Llama-S (unseen) & 0.0098 & 0.0066 & 0.0195 & 0.0097 & 0.0047 & 0.0032 & 0.0062 & 0.0038 &   0.0147 & 0.0108 & 0.0202 & 0.0126\\
        OpenP5-Llama-C (unseen) & 0.0005 & 0.0003 & 0.0015 & 0.0006 & 0.0003 & 0.0001 & 0.0004 & 0.0002 & 0.0037 & 0.0028 & 0.0037 & 0.0028\\
        SP5-Llama-R (unseen) & 0.0018 & 0.0009 & 0.0030 & 0.0013 & 0.0006 & 0.0003 & 0.0010 & 0.0004 & 0.0018 & 0.0009 & 0.0046 & 0.0019\\
        SP5-Llama-S (unseen) & 0.0048 & 0.0030 & 0.0098 & 0.0046 & 0.0011 & 0.0006 & 0.0014 & 0.0007 & 0.0018 & 0.0010 & 0.0037 & 0.0015\\
        SP5-Llama-C (unseen) & 0.0022 & 0.0011 & 0.0043 & 0.0018 & 0.0001 & 0.0001 & 0.0006 & 0.0002 & 0.0001 & 0.0001 & 0.0009 & 0.0003\\
        \bottomrule
    \end{tabular}
    \caption{Performance results on straighforward recommendation task. R, S, C represent three item indexing.}
    \label{tab:straighforward}
    \vspace{-20pt}
\end{table*}

\section{Experiments}
\label{sec:experiments}



\subsection{Datasets and Baselines}
We have introduced the dataset collection in Section \ref{sec:dataprocessing}. 
Due to the space limitation, we present experimental results on three datasets: Movielens-1M, Amazon Beauty, LastFM.
The remaining results can be accessed via our GitHub repository\footnote{https://github.com/agiresearch/OpenP5}.
To demonstrate the superior performance of the OpenP5 platform, we gather a collection of representative approaches for different downstream tasks.

\textbf{Sequential Recommendation.} Since our OpenP5 only uses user interaction information for prediction, for fair comparision, we adopt several prominent sequential recommendation baselines that also use user interaction information only.  We introduce the baseline models for sequential recommendation as follows.
\begin{itemize}[leftmargin=*]
    \item \textbf{Caser} \cite{tang2018personalized} views sequential recommendation as a Markov Chain and employ Convolutional Neural Networks (CNNs) to model users.
    \item  \textbf{HGN} \cite{ma2019hierarchical} uses hierarchical gating networks to learn user behaviors from both long-term and short-term perspectives.
    \item \textbf{GRU4Rec} \cite{hidasi2016session} leverages Gated Recurrent Units (GRU) \cite{cho2014learning} to model user interaction sequences.
    \item \textbf{Bert4Rec} \cite{sun2019bert4rec} utilizes BERT-style masked language modeling \cite{devlin2019bert} to learn a bidirectional representation for sequential recommendation.
    \item \textbf{FDSA} \cite{zhang2019feature} models feature sequences with a self-attention module.
    \item \textbf{SASRec} \cite{kang2018self} deploys a self-attention mechanism within a sequential recommendation model.
\end{itemize}

\textbf{Straightforward Recommendation.} We utilize three existing methods as our baselines for straightforward recommendation task. 
\begin{itemize}[leftmargin=*]
    \item \textbf{BPR-MF} \cite{rendle2009bpr} leverages matrix factorization with the pairwise Bayesian Personalized Ranking (BPR) loss.
    \item \textbf{BPR-MLP} \cite{cheng2016wide} utilize MLP to model users and items.
    \item \textbf{SimpleX} \cite{mao2021simplex} leverages cosine contrastive loss (CCL) in collaborative filtering for recommendation, which is a very strong baseline that beats many graph-based recommendation models.
\end{itemize} 

\subsection{Implementation Details}\label{sec:implementation}

Following the P5 framework \cite{p5}, our implementation is grounded in the T5 model \cite{raffel2020exploring} and LLaMA-2 \cite{touvron2023llama} model. The T5 backbone is trained on full parameters, while LLaMA-2 is trained using LoRA \cite{hu2021lora}. 
Notably, we randomly initialize the embedding of number-related tokens in the pre-trained checkpoint. This is predicated on the fact that while these embeddings encapsulate semantic similarity amongst tokens in the pre-training phase, such semantic patterns may not carry over to recommendation tasks when items are indexed with numerical identifiers. For prompt during the training, we select 10 prompts for each task, reserving one to evaluate zero-shot generalization. 
To alleviate the potential forgetting problem, we employ a training regimen wherein batches from different tasks are alternated. For SP5, addressing the data imbalance and potential forgetting problem is important. Hence, we alternate batches derived from different datasets and tasks. For smaller datasets, we repeat iterations until the completion of the largest dataset to ensure a balanced training process.

\subsection{Results Analysis}

The performance metrics for the sequential recommendation task and the straightforward recommendation task are presented in Table \ref{tab:sequential} and Table \ref{tab:straighforward} respectively. Specifically, we use the top-$k$ Hit Ratio (HR@$k$) and Normalized Discounted Cumulative Gain (NDCG@$k$) to evaluate performance, providing the results for HR@5,10, and NDCG@5,10. The best result for each metric is highlighted in bold, while the second-best result is underlined. 

From the recommendation performance shown in Table \ref{tab:sequential} and Table \ref{tab:straighforward}, we can observe that generative recommendation is capable of achieving the best performance in most cases compared with baselines but highly depends on the pre-trained backbone models and item indexing methods. Comparing T5-based OpenP5 and Llama-based OpenP5, the performance with T5 backbone is better than Llama backbone in most cases. This is potentially due to the large number of parameters in Llama backbone, which leads to underfitting with sparse recommendation data. For T5-based OpenP5, a comparison of the three item indexing methods reveals the anticipated lower performance of the random indexing method, with the sequential indexing method trailing slightly behind the collaborative indexing method. Conversely, the Llama-based OpenP5 model does not yield a clear preference for any indexing method, potentially attributable to the effects of sparse data on its larger parameter space. This shows future avenues for developing LLM-based generative recommendation models that are both effective and parameter-efficient.


\section{Conclusion and Future Work}\label{sec:conclusion}
In this paper, we provide the OpenP5 platform as a resouce to facilitate the developing, training, and evaluating large language model based recommender systems. We consider the implementation on four perspectives: backbone models,  downstream tasks, recommendation datasets, and item indexing methods. The platform serves as a continous effort to develop and evaluate foundation models for recommendation and helps the community to advance further on this direction with future innovations.
In the future, we will consider incorporating more item indexing methods, more foundation model training and inference paradigms, more data modalities, and more backbone LLMs into the platform.

\setcounter{subsection}{0}
\renewcommand{\thesubsection}{\Alph{subsection}}

\section*{Appendix}
In this appendix, we provide the full list of the personalized prompts for both downstream tasks.

\subsection{Sequential Recommendation}

\noindent\textbf{Prompt Seen: A1}\\
\texttt{Input Template: Considering \{\textcolor{blue}{dataset}\} user\_\{\textcolor{blue}{user\_id}\} has interacted with \{\textcolor{blue}{dataset}\} items \{\textcolor{blue}{history}\} . What is the next recommendation for the user ?\\
Target Template: \{\textcolor{blue}{dataset}\} \{\textcolor{blue}{target}\}
}


\noindent\textbf{Prompt Seen: A2}\\
\texttt{Input Template: Here is the purchase history of \{\textcolor{blue}{dataset}\} user\_\{\textcolor{blue}{user\_id}\} : \{\textcolor{blue}{dataset}\} item \{\textcolor{blue}{history}\} . I wonder what is the next recommended item for the user .\\
Target Template: \{\textcolor{blue}{dataset}\} \{\textcolor{blue}{target}\}
}

\noindent\textbf{Prompt Seen: A3}\\
\texttt{Input Template: \{\textcolor{blue}{dataset}\} user\_\{\textcolor{blue}{user\_id}\} has purchased \{\textcolor{blue}{dataset}\} items \{\textcolor{blue}{history}\} , predict next possible item to be bought by the user ? \\
Target Template: \{\textcolor{blue}{dataset}\} \{\textcolor{blue}{target}\}
}

\noindent\textbf{Prompt Seen: A4}\\
\texttt{Input Template: I find the purchase list of \{\textcolor{blue}{dataset}\} user\_\{\textcolor{blue}{user\_id}\} : \{\textcolor{blue}{dataset}\} items \{\textcolor{blue}{history}\} , I wonder what other itmes does the user need . Can you help me decide ?\\
Target Template: \{\textcolor{blue}{dataset}\} \{\textcolor{blue}{target}\}
}

\noindent\textbf{Prompt Seen: A5}\\
\texttt{Input Template: According to what items \{\textcolor{blue}{dataset}\} \\user\_\{\textcolor{blue}{user\_id}\} has purchased : \{\textcolor{blue}{dataset}\} items \{\textcolor{blue}{history}\} , Can you recommend another item to the user ?\\
Target Template: \{\textcolor{blue}{dataset}\} \{\textcolor{blue}{target}\}
}

\noindent\textbf{Prompt Seen: A6}\\
\texttt{Input Template: What would \{\textcolor{blue}{dataset}\} user\_\{\textcolor{blue}{user\_id}\} be likely to purchase next after buying \{\textcolor{blue}{dataset}\} items \{\textcolor{blue}{history}\} ?\\
Target Template: \{\textcolor{blue}{dataset}\} \{\textcolor{blue}{target}\}
}

\noindent\textbf{Prompt Seen: A7}\\
\texttt{Input Template: By analyzing the \{\textcolor{blue}{dataset}\} user\_\{\textcolor{blue}{user\_id}\} 's purchase of \{\textcolor{blue}{dataset}\} items \{\textcolor{blue}{history}\} , what is the next item expected to be bought ?\\
Target Template: \{\textcolor{blue}{dataset}\} \{\textcolor{blue}{target}\}
}

\noindent\textbf{Prompt Seen: A8}\\
\texttt{Input Template: Can you recommend the next item for \{\textcolor{blue}{dataset}\} user\_\{\textcolor{blue}{user\_id}\} , given the user 's purchase of \{\textcolor{blue}{dataset}\} items \{\textcolor{blue}{history}\} ?\\
Target Template: \{\textcolor{blue}{dataset}\} \{\textcolor{blue}{target}\}
}

\noindent\textbf{Prompt Seen: A9}\\
\texttt{Input Template: After buying \{\textcolor{blue}{dataset}\} items \{\textcolor{blue}{history}\} , what is the next item that could be recommended for \{\textcolor{blue}{dataset}\} user\_\{\textcolor{blue}{user\_id}\} ?\\
Target Template: \{\textcolor{blue}{dataset}\} \{\textcolor{blue}{target}\}
}

\noindent\textbf{Prompt Seen: A10}\\
\texttt{Input Template: The \{\textcolor{blue}{dataset}\} user\_\{\textcolor{blue}{user\_id}\} has bought items : \{\textcolor{blue}{dataset}\} items \{\textcolor{blue}{history}\} , What else do you think is necessary for the user ?\\
Target Template: \{\textcolor{blue}{dataset}\} \{\textcolor{blue}{target}\}
}

\noindent\textbf{Prompt Unseen: A11}\\
\texttt{Input Template: What is the top recommended item for \{\textcolor{blue}{dataset}\} user\_\{\textcolor{blue}{user\_id}\} who interacted with \{\textcolor{blue}{dataset}\} item \{\textcolor{blue}{history}\} ?\\
Target Template: \{\textcolor{blue}{dataset}\} \{\textcolor{blue}{target}\}
}

\subsection{Straightforward Recommendation}

\noindent\textbf{Prompt Seen: B1}\\
\texttt{Input Template: What should we recommend for \{\textcolor{blue}{dataset}\} user\_\{\textcolor{blue}{user\_id}\} ?\\
Target Template: \{\textcolor{blue}{dataset}\} \{\textcolor{blue}{target}\}
}

\noindent\textbf{Prompt Seen: B2}\\
\texttt{Input Template: \{\textcolor{blue}{dataset}\} user\_\{\textcolor{blue}{user\_id}\} is looking for some items . Do you have any recommendations ?\\
Target Template: \{\textcolor{blue}{dataset}\} \{\textcolor{blue}{target}\}
}

\noindent\textbf{Prompt Seen: B3}\\
\texttt{Input Template: Do you have any suggested items for {dataset} user\_\{\textcolor{blue}{user\_id}\} ?\\
Target Template: \{\textcolor{blue}{dataset}\} \{\textcolor{blue}{target}\}
}

\noindent\textbf{Prompt Seen: B4}\\
\texttt{Input Template: Which recommendation should we provide to \{\textcolor{blue}{dataset}\} user\_\{\textcolor{blue}{user\_id}\} ?\\
Target Template: \{\textcolor{blue}{dataset}\} \{\textcolor{blue}{target}\}
}

\noindent\textbf{Prompt Seen: B5}\\
\texttt{Input Template: How can we assist \{\textcolor{blue}{dataset}\} user\_\{\textcolor{blue}{user\_id}\} with a recommendation ?\\
Target Template: \{\textcolor{blue}{dataset}\} \{\textcolor{blue}{target}\}
}

\noindent\textbf{Prompt Seen: B6}\\
\texttt{Input Template: What would be a suitable recommendation for \{\textcolor{blue}{dataset}\} user\_\{\textcolor{blue}{user\_id}\} ?\\
Target Template: \{\textcolor{blue}{dataset}\} \{\textcolor{blue}{target}\}
}

\noindent\textbf{Prompt Seen: B7}\\
\texttt{Input Template: What would be a helpful recommendation for \{\textcolor{blue}{dataset}\} user\_\{\textcolor{blue}{user\_id}\} ?\\
Target Template: \{\textcolor{blue}{dataset}\} \{\textcolor{blue}{target}\}
}

\noindent\textbf{Prompt Seen: B8}\\
\texttt{Input Template: Can you recommend an item for \{\textcolor{blue}{dataset}\} user\_\{\textcolor{blue}{user\_id}\} ?\\
Target Template: \{\textcolor{blue}{dataset}\} \{\textcolor{blue}{target}\}
}

\noindent\textbf{Prompt Seen: B9}\\
\texttt{Input Template: Based on \{\textcolor{blue}{dataset}\} user\_\{\textcolor{blue}{user\_id}\} 's interests and requirements , what item would you suggest to try ?\\
Target Template: \{\textcolor{blue}{dataset}\} \{\textcolor{blue}{target}\}
}

\noindent\textbf{Prompt Seen: B10}\\
\texttt{Input Template: For \{\textcolor{blue}{dataset}\} user\_\{\textcolor{blue}{user\_id}\} , what item stands out as a top recommendation that they should consider ?\\
Target Template: \{\textcolor{blue}{dataset}\} \{\textcolor{blue}{target}\}
}

\noindent\textbf{Prompt Unseen: B11}\\
\texttt{Input Template: What is the top recommendation for \{\textcolor{blue}{dataset}\} user\_\{\textcolor{blue}{user\_id}\} ?\\
Target Template: \{\textcolor{blue}{dataset}\} \{\textcolor{blue}{target}\}
}

\bibliographystyle{ACM-Reference-Format}
\bibliography{ref}


\end{document}